# Stratospheric temperature anomalies as imprints from the dark Universe

K. Zioutas[1)*], A. Argiriou[1)], H. Fischer[2)&], S. Hofmann[3)],
M. Maroudas[1)], A. Pappa[1)], Y.K. Semertzidis[4)]

1) Physics Department, University of Patras, 26504 Patras, Greece.
2) Physikalisches Institut, Albert-Ludwigs-Universität Freiburg, 79104 Freiburg, Germany.
3) Munich, Germany
4) Center for Axion and Precision Physics Research, Institute for Basic Science, and, Department of Physics, KAIST, Daejeon, 34141, Republic of Korea,

  *) Zioutas@cern.ch    &) Horst.Fischer@cern.ch

**ABSTRACT**
The manifestation of the dark Universe begun with unexpected large-scale astronomical observations. Here we are investigating the possible origin of small-scale anomalies, like that of the annually observed temperature anomalies in the stratosphere (38.5 – 47.5 km). Unexpectedly within known physics, we observe a planetary relationship of the daily stratospheric temperature distribution. Interestingly, its spectral shape does not match concurrent solar activity (F10.7 line), or Sun's EUV emission, whose impact on the atmosphere is unequivocal; this different behaviour points at an additional energy source of exo-solar origin. A viable concept behind such observations is based on possible gravitational focusing by the Sun and its planets towards the Earth of low-speed invisible (streaming) matter. When the Sun-Earth direction aligns with an invisible stream, its influx towards the Earth gets temporally enhanced. We denote generic constituents from the dark Universe as "invisible matter", in order to distinguish them from ordinary dark matter candidates like axions or WIMPs, which cannot have any noticeable impact. Moreover, the observed peaking planetary relations exclude on their own any conventional explanation, be it due to any remote planetary interaction, or, intrinsic to the atmosphere. Only a *somehow* "strongly" interacting invisible streaming matter with the little screened upper stratosphere ($\rho_{overhead} \approx 1$ gr/cm$^2$) can be behind the occasionally observed temperature increases. We also estimate an associated energy deposition O(~W/m$^2$), which is variable over the 11-years solar cycle. For the widely assumed picture of a quasi not-interacting dark Universe, this new exo-solar energy is enormous. Noticeably, our observationally derived conclusions are not in conflict with the null results of underground dark matter experiments, given that a similar planetary relationship is not observed even underneath the stratosphere (16-31 km). Interestingly, the atmosphere is uninterruptedly monitored since decades. Therefore, it can serve also parasitically as a novel (low threshold) detector for the dark Universe, with built-in spatiotemporal resolution and the Sun acting temporally as signal amplifier. Known phenomena (e.g., NAO, QBO and ENSO) influencing the general atmospheric circulation do not interfere with this work, since they occur geographically elsewhere, and, they have different periodicities. In future, analyzing more observations, for example, from the anomalous ionosphere, or, the transient sudden stratospheric warmings, the nature of the assumed "invisible streams" could be deciphered.

# 1. Introduction

The observation of an anomalous atmospheric ionization (1912) resulted to the accidental discovery of cosmic rays [1]. Similarly, here we explore instead the origin of temperature anomalies in the stratosphere [2,3] (altitude ≈ 40 km). Since the Earth's atmosphere is a complex medium, the challenging question is whether such anomalies are intrinsic to the atmosphere, or, they are triggered externally. Some previous observations are even consistent with a propagating "signature" from the stratosphere (altitude ≈ 10-50 km) downwards [3,4,5]. However, if the stratospheric temperature does not follow the simultaneously measured variable solar activity, which strongly affects our atmosphere, such anomalies should be of exo-solar origin, i.e., even more intriguing.

Interestingly, a variety of previous atmospheric observations have shown that the Earth's (upper) atmosphere senses *somehow* the 11-years solar cycle ([6,7,8]), which, within known physics, remains one of the oldest and biggest unsolved mysteries in solar physics [9,10]. Notably, the orbital period of Jupiter (11.86 years), or, the synod Jupiter-Earth-Venus (11.01 years) are probably not symptomatically close to the 11-years periodicity, while, following ref's [10,11], the solar cycle is planetary driven.

Furthermore, because of solar cycle's manifestation in a plethora of phenomena, one is used to accept it as something obvious. However, there is no conventional explanation for a remote planetary interaction with Earth's atmosphere, neither by gravitational nor by any other forces, since they are too weak [10]. In addition, the observation of a peaking planetary relationship excludes on its own any remote interactions, since their strength changes smoothly during an orbit [11]. Thus, in order to identify the origin of some atmospheric signature (in particular with a possible 11-years rhythm), the search for a planetary relationship is inevitable.

The driving idea behind this study is based on the gravitational focusing by the Sun and its planets of low-speed invisible (streaming) matter. Whatever its ultimate properties, it must interact *somehow* with the upper atmosphere, in order to be able to cause the observed anomalous stratospheric behaviour. We refer to generic dark matter constituents as "*invisible massive matter*" [11], in order to distinguish them from the widely addressed dark matter candidates like axions or WIMPs, which cannot have any noticeable atmospheric impact. Encouragingly, recent work discusses potential constituents from the dark sector [12,13,14,15] having a large cross section with normal matter.

## 2. The Stratospheric Temperature Data

The atmospheric temperature data used in the present study come from the ERA – Interim global reanalysis dataset [16]. The reanalysis data are created via a data assimilation scheme and models which ingest all available observations every 6-12 hours over the period being analyzed. Under the current state-of-the-art, approximately 7-9 million observations are ingested at each time step. The specific data set used here covers the period from January 1st, 1986 until present. The spatial resolution of the data set is approximately 0.125° × 0.125°, on 60 vertical levels from the surface up to the 0.1 hPa isobaric level (height ≈ 80 km) [17]. Measurements used for the assimilation are derived mostly from satellite measurements but also from radiosondes, pilot balloons, aircraft, and wind profilers. The number of available data for these sources are practically constant during this period, except for the aircraft reports whose numbers increased significantly after 1998 [16].

For this work we used the atmospheric temperature data at the grid points (42.5°N, 13.5°E with size equal to about 10.7 km (direction West - East) by 13.8 km (direction North - South) and at the isobaric levels of 3 hPa, 2 hPa and 1 hPa, the approximate altitude of which is 38.5 km, 42.5 km and 47.5 km, respectively, in relation to a standard atmosphere. For each isobaric level, two data sets were retrieved, one at 00:00 and another at 12:00 UTC, provided by the European Center for Medium-range Weather Forecasts (ECMWF) [18]. The daily values used for our analysis are the corresponding arithmetic averages [(00:00+12:00)/2]. Then, the average of the 3 isobaric layers is calculated. The overall uncertainty of an initial single temperature measurement is estimated to be about ±0.5 K [19].

## 3. Analysis – Results

In this work, we have chosen a grid point close to INFN Laboratori Nazionali del Gran Sasso (LNGS) / Italy, in order to obtain collocated data sets describing the state of the atmosphere overhead. The reason for choosing this location and the time span of 10 years, it is the recently published results by the Borexino collaboration [20], showing annual and long-term modulation of the cosmic muon flux during the data taken period 17/5/2007-17/5/2017. The measured relative amplitude of (1.36±0.04)% is much less than the stratospheric temperature modulation observed in this work, which show also a strong peak around December-January (see below). Figure 1 shows the time dependent mean daily upper stratospheric temperature (1986 - 2018). Throughout this work we have analysed only the data available from the utmost upper stratosphere (3, 2 and 1 hPa). The temperature signatures appearing in the upper stratosphere, appear also at lower altitudes, though with decreased amplitude (not shown here). Around the months of December-January (or, around 70º to 120º Earth's heliocentric longitudes) striking temperature excursions appear each year on top of the rather smooth seasonal variations. Previous published plots also show temperature anomalies in the stratosphere [21] confirming qualitatively the derived time spectrum in this work (Figure 1). The purpose of this investigation is to find out the origin of such a burst-like heating of the upper stratosphere around December-January each year.

In order to establish a planetary effect, we project the mean daily stratospheric temperature on the associated planetary longitudinal coordinates of that day. If an orbiting period is short compared to the data taken period, this is equivalent of repeating one measurement several times. Therefore, the inner planets may be better suited to exclude random or systematic effects, as well as to establish a new planetary effect. In this work the eccentricity related modulation is factored-out, and therefore, in the absense of any planetary connection, the stratospheric temperature spectra should be smoothly distributed following seasonal influence (but without peaks).

In addition, in order to exclude that the Sun's irradiation causes the temperature anomalies, additional distributions of potential interest are derived under identical conditions using the daily measured solar activity proxy F10.7 radio line (≈2.8 GHz), and, the solar EUV emission, whose atmospheric impact is unequivocal. Since the solar activity is also planetary dependent [11], only if solar and stratospheric spectral shapes are dissimilar, this would point at an as yet unnoticed exo-solar stratospheric link. Taken into account the prevailing gravitational focusing effect by the Sun, such a link might point also towards but beyond the Sun.

Figure 2 shows schematically the gravitational focusing by the Sun towards the Earth of an incident low speed cosmic stream. Since the orbiting planets affect periodically the dominating Sun's gravitational field, they should give rise to a planetary imprint (=relationship), if an atmospheric signature is caused by a gravitationally focused stream.

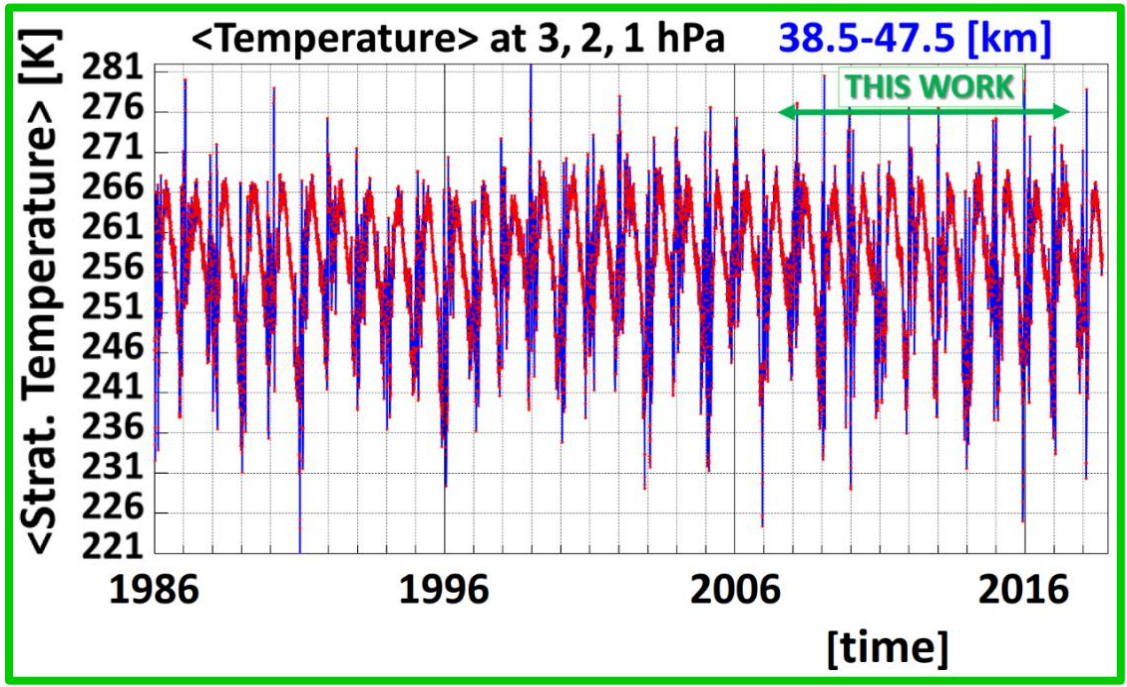

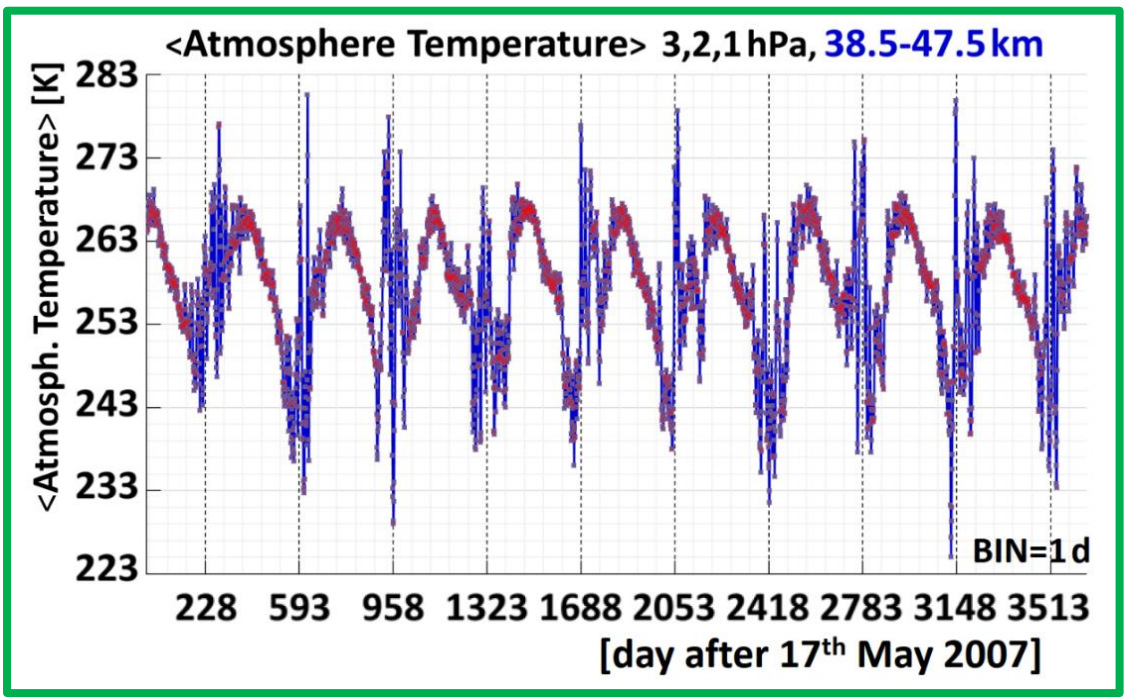

**Figure 1** (Top) Time dependence of the mean daily stratospheric temperature [(00:00+12:00)/2] at 3, 2, 1 hPa (altitude ≈ 38.5, 42.5, 47.5 km), 42.5°N / 13.5°E and for the period 1986-2018. The period analysed in this work is indicated and it is also shown expanded (Bottom). The vertical dashed lines are year boundaries: 1st January of 2008 … 2017. The error bar of each point is equal to 0.5 K [19].

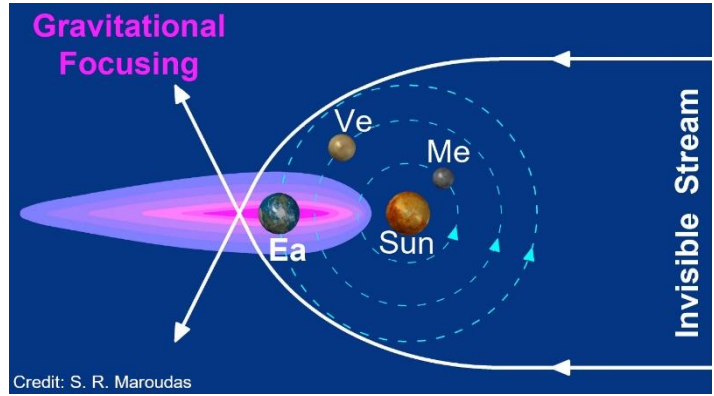

**Figure 2**  Schematic view of the flow of a putative slow speed invisible matter stream, which is gravitationally focused by the Sun towards the Earth. The orbiting planets disturb periodically the dominating Sun's gravitational focusing effect. For orientation purposes, in this specific Earth-Sun configuration around the 18$^{th}$ December (Longitude ≈ 86.5º), the assumed invisible stream comes (within 5.5º) from the direction of the Galactic Center (Longitude ≈ 266.5º). Note that the Sun is orbiting around the Galactic Center moving towards the constellation of Cygnus, i.e., upwards, perpendicularly to the direction of the invisible stream. This simplified gravitational focusing presentation is based on ref. [22].

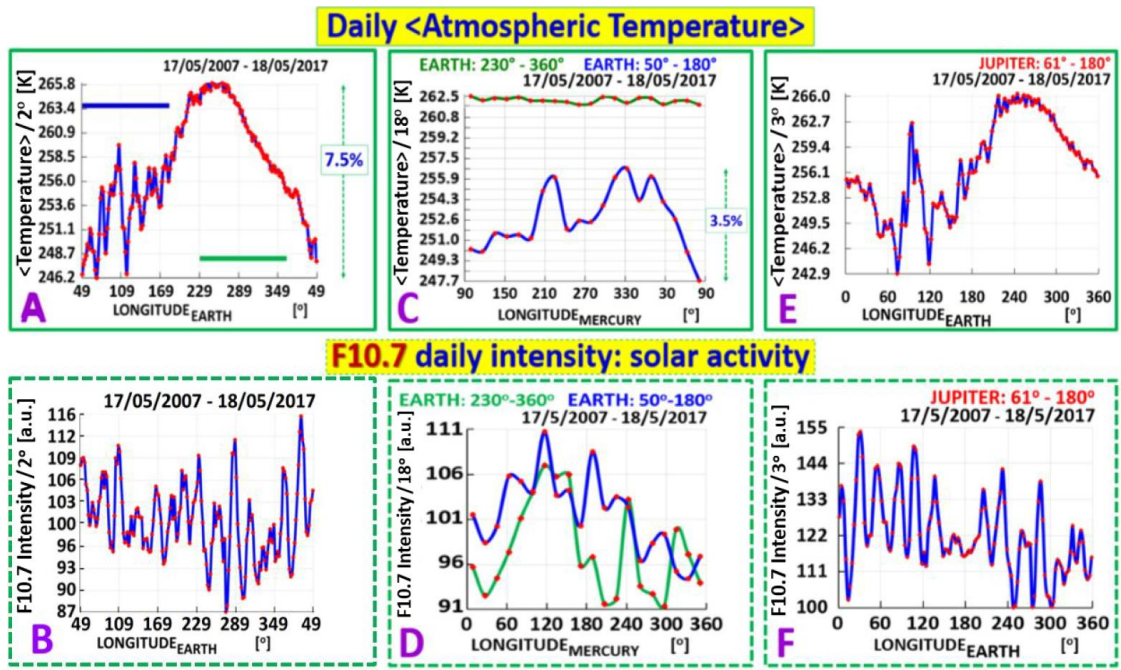

**Figure 3**     (Top) Planetary longitudinal distributions for the mean daily temperature [(00:00+12:00)/2] of the upper stratospheric layers at 3,2,1 hPa (altitude ≈ 38.5, 42.5, 47.5 km), 42.5°N / 13.5°E and for the period 2007-2017. The blue and green bars in (A) give the Earth's orbital constraints used in (C), which shows planetary relationship for Mercury's reference frame only if the Earth propagates in the heliocentric orbital arcs 50º to 180º.  Also the plot in (E) shows a clear planetary relationship.
(Bottom, inside dashed frames) The spectra (B), (D) and (F) show the corresponding longitudinal distributions of the F10.7 solar line (≈ 2.8 GHz), which is a proxy for the solar activity. A comparison between (A) and (B), (C) and (D), and, (E) and (F) shows no similarity between each pair. This excludes that the solar radiation is solely at the origin of the upper pannels.

Figure 3 (upper pannels) shows the first analysis results, where the mean daily stratospheric temperature is shown as a function of the corresponding planetary longitude. For comparison, the lower pannels give the corresponding distributions of the solar proxy (10.7 cm line) applying the same conditions in the analysis. The missing similarity excludes that the solar radiation is solely at the origin of the upper pannels, pointing hence to an (additional) exo-solar impact on the upper stratosphere.

It is worth mentioning that every planet follows its own "clock", smearing out any effect being not related to its own orbit. Therefore, in order to strengthen a potential planetary effect which is observed, e.g., in Earth's frame of reference, we reconstruct similar plots using other planetary reference frames, or combining one or more planets.

Thus, the spectrum in Figure 3A shows the averaged ten years temperature distribution of Figure 1 (bottom), projected to one year as a function of the Earth's longitudinal coordinates with 2° binning. The annually occuring temperature excursions, around December-January (Figure 1), appear as a peak in the Earth's spectra around ~100° (~1$^{st}$ January). In fact, the alignment Earth –Sun-Galactic Center, with the Earth's longitude being about 87°, coincides with the onset of this peak (see also Figure 8).

In this work we focus on the peaking first half of the Earth's spectrum, aiming to unravel the origin of the temporally heating-up of the upper stratosphere, as the second half is rather smooth. For this purpose, a number of spectra are presented, in order to follow the planetary relationship, whose significance becomes stronger when more spectra are seen combined.

For example, Figure 3C shows two Mercury spectra, requiring the Earth to be in two opposite orbital heliocentric arcs as it is shown (color coded) in Figure 3A and 3C. Interestingly, the lower blue curve in Figure 3C, which covers the peaking period of the Earth spectrum, shows an amplitude (maximum - minimum) of about 3.5%, which is comparable with that of the peak on the left of Figure 3A. I.e., in Mercury's reference frame smearing does not occur, even though Mercury performed 44 orbits during the 10 years of the analyzed data. Even more, the 2 or 3 sub-peaks strengthen further the planetary imprint in the blue spectrum of Figure 3C. However, it goes beyond the scope of the present work to identify such, even significant, spectral details. The comparison with the remarkably smooth green upper curve in Figure 3C is important for three reasons: a) Here, the 44 Mercury orbits do smear out the original decreasing temperature distribution of Earth's spectrum (Figure 3A, green line), becoming flat with an amplitude (maximum-minimum) of only 0.3%, b) Its strikingly small fluctuations show the expcted fluctuations of the blue lower curve, in spite of the diminishing error bars per bin. This means that when the Earth propagates 180° opposite, i.e., between 230° and 360°, in Mercury's reference frame a planetary dependence does not exist, and c) By using the same temperature data, the upper green curve serves as a very good "simulation" for the lower blue curve. In other words, the green line shows the null effect, since any unforeseen fluctuations should appear similarly in both spectra of Figure 3C.

Further, assuming that stratospheric planetary effects do not occur, the Earth spectrum (Figure 3A) should be replicated by requiring, e.g., the outer planet Jupiter was orbiting between 61° and 180° (Figure 3E). The selected longitude range corresponds to uninterupted 4 Earth orbits, in order to exclude any systematics when comparing Figure 3E with 3A. In fact, the peak around 70° to 120° appears in both spectra but with a different amplitude relative to the height of the wide maximum around 250°. This simple comparison is on its own an indication of planetary involvement.

Therefore, seen combined the 3 upper plots of Figure 3, they prove already that a planetary effect must be at work for the upper stratospheric temperature [2,3]. In addition, the dissimilarities between Figures 3A, 3C and 3E (stratospheric temperature) and 3B, 3D and 3F (solar activity), respectively, are pointing in addition at an exo-solar origin of the dynamic upper stratosphere. Remarkably, this external source is influencing the stratosphere at a similar level as the bright Sun.

Because of the importance of claiming such an as yet overlooked stratospheric signature, we add more supporting evidence, since each additional plot improves the confidence in the whole analysis. For example, both Venus spectra in Figure 4C confirm those of Mercury (Figure 3C); again, the lower spectrum has a different spectral shape and an even larger amplitude (~4.8%). In addition, applying double planetary constraints, we obtain Figure 4D. The striking peak around 100° is one of the strongest observed signature, strengthening the claim of an otherwise unexpcted planetary relationship of the dynamic upper stratosphere.

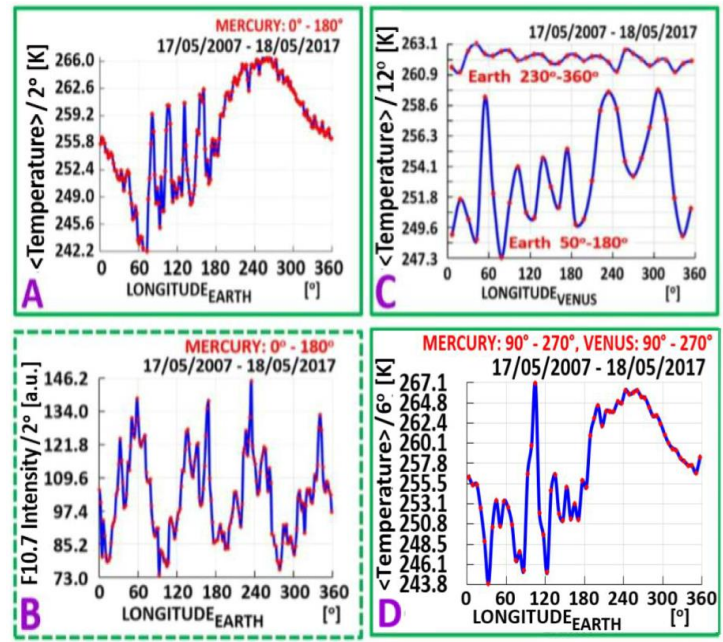

**Figure 4:** Planetary longitudinal distributions for the upper stratospheric temperature (A), and, inside the dashed frame, the solar activity given by the F10.7 solar line ($\approx$ 2.8 GHz) (B). The so much dissimilar spectral shapes exclude conventional solar radiation being the main cause of the upper atmospheric temperature distribution. The longitudinal distribution of the mean daily atmospheric temperature at 3,2,1 hPa (altitude$\approx$38.5–47.5 km) and location at 42.5°N / 13.5°E is given for Earth (A,D) and Venus (C), respectively. This is a further proof that a conventionally unexpected planetary relationship is at work in the dynamic upper atmosphere, pointing also at an exo-solar origin (see Figure 3).

In Figure 5 we compare more spectral shapes of the stratospheric temperature with the solar proxy (F10.7 line), using the frame of reference of Mercury, Venus and Mars, without applying any planetary constraint (for the Earth see Figure 3(A,B)). Again, the dissimilarity of the spectral shapes between the mean temperature and the F10.7 radio line is apparent, which supports the aforementioned exo-solar scenario. In addition, by using these different reference frames, Figure 5 (upper pannels) confirms the underlying planetary relationship for the upper stratospheric temperatures.

Figure 6 gives more spectra, which allow to follow the behaviour of the temperature peak around 100° in Earth longitudes for a number of planetary configurations. Moreover, we also analyzed the stratospheric and F10.7 data in a much longer time period (1986-2018), which further validate the ten years study (see below Figure 7 and 8).

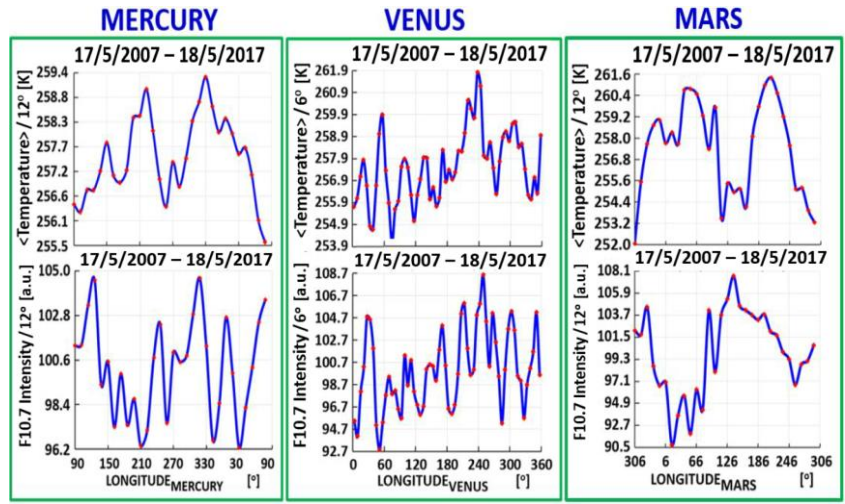

**Figure 5** The mean daily temperature distribution measured at 3,2,1 hPa (altitude ≈38.5–47.5 km) and at the location 42.5°N / 13.5°E as a function of Mercury, Venus and Mars longitude is given in the upper spectra. For comparison, the lower spectra show the solar activity derived from the daily intensity of the F10.7 solar line for the same time interval (2007-2017). The error bars are barely seen. For Mercury and Mars the spectral shapes are clearly different, while for Venus this is less pronounced. Along with the highly different Earth's spectral shapes (see Figure 3(A,B), E&F and Figure 4 A&B), the comparison temperature vs. F10.7 line in this Figure further excludes known solar radiation being the only driving source behind the stratospheric temperature distributions.

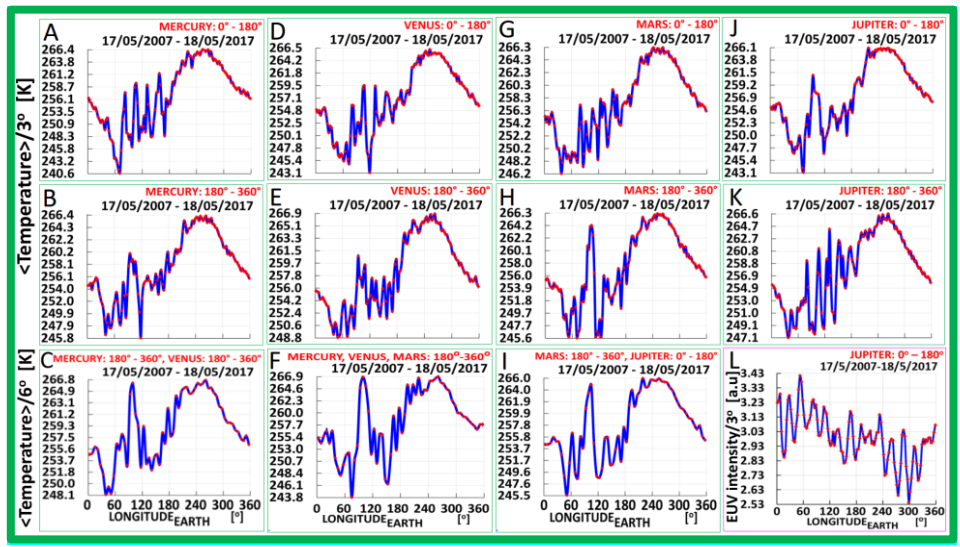

**Figure 6** (Top and middle) Additional upper stratospheric mean daily temperature spectra as a function of Earth's longitude combined with different planetary constraints (0º-180º) *vs.* (180º-360º). (bottom) combining multiple planetary constraints, which are applied in B&E, B&E&H and H&J, we arrive at stronger peaking distributions in C, F and I , respectively. The peak around 100º appears with a better signal-to-noise ratio.

The spectral distribution of the daily measured solar EUV intensity for the same conditions as in J, is given in L. The dissimilarity with the corresponding temperature spectrum in J is remarkable. This excludes conventional solar EUV to be behind the peak around 100º, favouring an exo-solar impact.

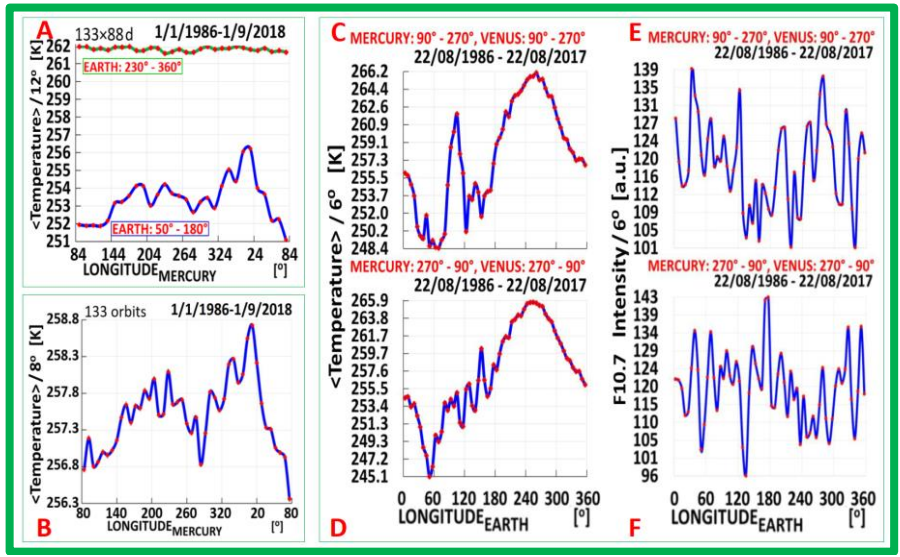

**Figure 7** The various spectra are from the upper stratosphere (3,2,1 hPa) and the measuring period is 1986-2018. The spectra A, B and C are similar in shape and correspond to Figure 3C, 5 (Mercury) and 4D, respectively. Spectra (E, F), obtained with the F10.7 cm line, are strikingly different from the upper stratospheric temperature distributions (C, D), excluding thus an involvement of the solar activity exclusively. Concerning the orbital constraints for Mercury and Venus, spectra (C) and (D) are complementary to each other (180° opposite), confirming the observed strongest planetary relationship of the peaking upper stratospheric temperature covering 31 years . Mercury and Earth performed 133 and 31 orbits, respectively.

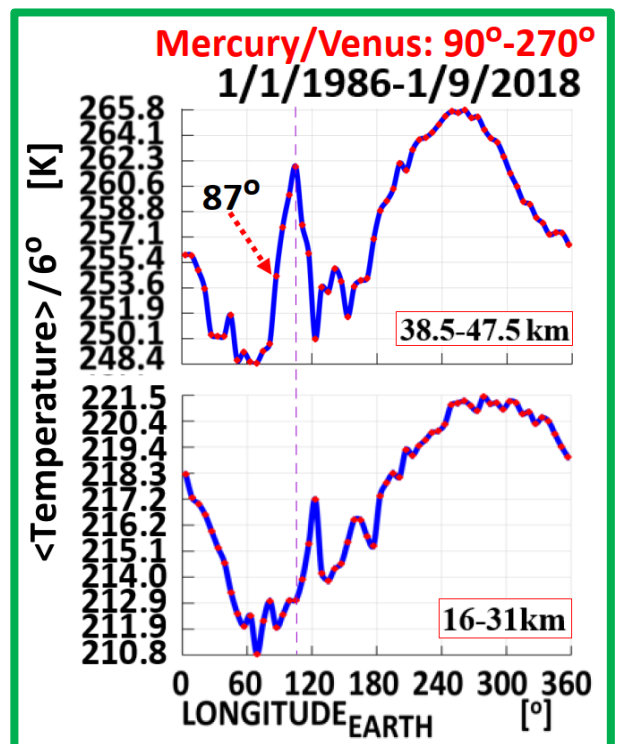

**Figure 8** A comparison between the mean temperature spectra of the upper stratosphere (top) and the lower stratosphere (bottom). The lower stratosphere (16-31 km) is the main Ozone layer, which is strongly affected by the solar UV. The striking difference between both spectra implies that the upper stratosphere (38.5-47.5 km) is marginally or even not affected at all by the solar UV. The position of the Galactic Center in this plot is at ~86.5°, and the upper stratosphere reaches its maximum temperature ~18 days later.

## 3.1 Systematics

In order to further exclude possible systematics behind the results of this work, we have divided the 10 years data taken period in consecutively 44×88 days sub-periods. The 88 days coincide with Mercury's orbital period. Figure 9 gives their sum as well as their product day-by-day. The initial maximum-to-minimum amplitude of the summed spectrum is 1.2%. If the observed distribution would arise randomly, then the 44 times multiplied spectrum should become even more randomly [11], which is apparently not the case. On the contrary, assuming that each of the 44 partial spectra are identical with an amplitude of 1.2% (ideal case), the 44 times multiplied partial spectra should appear with a similar shape, but with an amplitude of 69% in the multiplication spectrum ($1.012^{44}$=1.69). While the shape remains, the observed amplitude is 43%. Given the statistical nature of such observations, this actually reasonable value indicates that the underlying mean temperature "fluctuations" match quite coherently Mercury's 44 orbits during the 10 years. This is an important cross checking for the advocated planetary relationship. This result has been confirmed using 10×1 year instead of 44×88 days (not shown).

In addition, the recovery of spectral shapes, using a 3× longer data taken period, improves the confidence to the whole analysis procedure (see, indicatively, caption of Figure 7). After all, Mercury and Earth performed, 133 and 31 orbits, instead of 44 and 10, respectively.

Further, the comparison, for example, of the upper stratosphere with the ozone layer underneath (see Figure 8), provides additionally an independent reasoning in favour of this work.

Finaly, combining all shown observations, it strengthens the evidence supporting the working hypothesis within the dark sector, i.e., of invisible streaming matter, while other interpretations do not fit-in.

## 3.11 Other atmospheric phenomena

The Earth's atmosphere is a complex, non-linear dynamic system where several phenomena take place at different spatiotemporal scales. In fact, due to threshold effects, the same energy deposition in the visible, UV or X-rays has quite different impact on the dynamic atmosphere. This applies also to the unknown exo-solar source of this work when calculating its energy deposition.

One atmospheric phenomenon of potential interest seems to be that of the sudden stratospheric warmings (SSWs).Though, these are rare events occurring rather randomly, and therefore, they do not interfere with the striking annually occurring stratospheric temperature anomalies (STAs). There exist also major SSWs being associated with rapid (a few days) and large temperature increase (~50K), occurring over the polar zone of the Northern Hemisphere at a frequency of about six events per decade [26]. Admittedly, SSWs deserve full attention following the reasoning of this work. We leave this for a future investigation.

Another STA is the Quasi Biennial Oscillation (QBO), manifested as downward propagating easterly and westerly wind regimes, with a variable period averaging approximately 28 months. The QBO modifies the temperature across the stratosphere of about 0.3-1 K [27]. The equatorial temperature anomalies associated with the QBO in the lower stratosphere are of the order of ±4 K, maximizing near the isobaric levels from 30 to 50 hPa (up to 7K) [28]. The QBOs however cannot explain the anomalies analyzed in this paper, since its period of occurrence is much larger than a year.

Major surface pressure differences, impacting the general circulation of the atmosphere like the North Atlantic Oscillation (NAO) or the variations such as the El Niño / Southern Oscillation (ENSO) (in the Pacific), can generate sea-surface temperature anomalies that trigger stationary Rossby waves in the Extratropics [29]. Both the NAO and the ENSO vary interannually. The NAO has no specific periodicity and the ENSO has an irregular periodicity that may range from two to seven years, but they do influence the general atmospheric circulation. Since they occur outside the geographic area chosen here, and, they have totally different periodicities, they do not interfere with this work, at least not directly.

Moreover, the near surface air temperature, the temperature of the troposphere and stratosphere vary spatiotemporally. The temperature of the stratosphere varies at different spatiotemporal scales due to the non-homogeneous distribution of the continents and the oceans around the globe, and also because of the variation of the impinging solar radiation; all the above are parameters affecting the general atmospheric circulation.

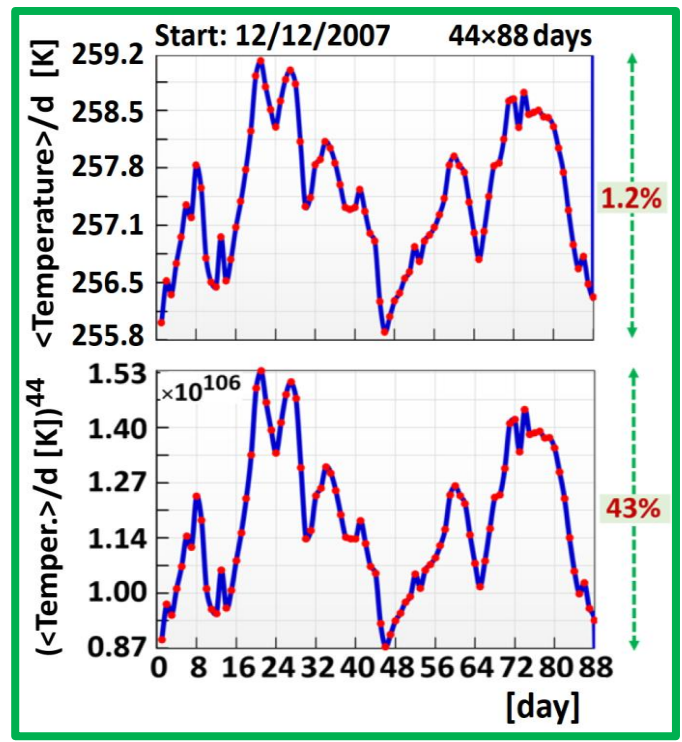

**Figure 9** (Top) The sum of 44×88 days consecutive daily mean temperature spectra during the 10 years measuring period. The maximum amplitude is 1.2%. (bottom) By multiplying all 44 partial spectra day-by-day, the derived spectrum has an amplitude of 43%, with 69% being the expected value in the ideal case.

### 3.2 Discussion

Here, we explore for the first time a possible planetary correlation of the daily stratospheric temperature. The driving idea behind this work is at the moment the only viable concept for the otherwise unexpectedly observed planetary relationship; namely, it is based on the gravitational focusing by the Sun and its planets towards the Earth's atmosphere of low-speed invisible (streaming) matter from the Milky Way or of cosmic origin.

Atmospheric oscillations like NAO, QBO and ENSO (see section 3.11) do influence the general atmospheric circulation. Since they occur outside the geographic area chosen here, and, they have totally different periodicities (e.g., from intraannual to ~7 years), they cannot be at the origin of the temperature anomalies in the upper stratosphere. The situation ends-up in a dead-end, as with known physics one has to also explain (peaking) planetary relationships derived in this work.

Whatever the ultimate properties of the invisible streams, they must interact *somehow* with the atmosphere, in order to cause the derived anomalous stratospheric behaviour. Interestingly, recently two different works ([12] and [13,14,15]) discuss potential dark matter constituents with a large cross section with normal matter.

The observed temperature anomalies inspired this work to exclude an origin intrinsic to the coupled system atmosphere / solar-irradiation (ASI). Therefore, we projected the daily temperature values into other planetary reference frames. Since remote planetary interactions are extremely tiny, any effect intrinsic to the system ASI should be smeared out, and this is what we observe. For example, this is best confirmed by using the same dataset of this work. Figure 3(C) shows the derived stratospheric temperature distributions in the Mercury's reference frame using two different time periods for each year: a) the transient temperature excursions (blue), and b) the smooth distributions (green). Remarkably, in contrast to the peaking periods, the smooth ones are almost smeared out as expected, in case no planetary relationship is present; and this in spite of the fact that the time dependent values used for the green line decrease with time as it is shown in Figure 3(A). Therefore, within Mercury's frame of reference, the green line almost perfectly simulates (!) periods of actually missing external impact to the system ASI. Noticeably, the same conclusion is derived by using a more than 3× longer period of data taken (see Figure 7(A)). In addition, Figure 4(C) confirms this behavior when Venus, instead of Mercury, is used. Thus, 3 different double spectra show a similar behavior.

It is worth mentioning, that in this kind of investigation a simulation is not easy, since the atmosphere itself is a not easy medium. Furthermore, because of the importance of the claim of this work, in order to improve its credibility and make the claim of the planetary scenario more robust, we have extended the data analysis arriving to a number of additional spectra. Among the diverse observations, the one given in Figure 8 *might be* the mostly synoptic one (see section 3.3).

<u>*In conclusion*</u>, based on the otherwise unexpected planetary relations, we argue in favour of the only viable explanation for the annual stratospheric temperature anomalies, namely that of streaming invisible massive particles being also exo-solar in origin. We stress that the observed peaking planetary longitudinal distributions (beyond Earth's alone) exclude on their own any remote planetary interactions, which are smooth over a planetary orbit.

### 3.3 Towards an exo-solar source

A question that arises is whether the observed planetary correlation of the upper stratosphere appears indirectly entirely due to the planetary dependence of the solar activity itself [11]. However, the apparent dissimilarity between solar and stratospheric spectral shapes (see Figure 3, 4 and 5) does not favour exclusively such an indirect scenario. On the contrary, it is rather pointing at an additional as yet invisible low speed streaming matter from outer space ("exo-solar"), which gets occasionally aligned with the Sun - Earth direction. Gravitational focusing by the Sun (and its planets) increases its flux at the focal plane [23], which can be at the position of the Earth. If one could trace back such a stream, it might appear to originate, e.g., from the Sun position in space, but its actual origin can be somewhere behind the Sun. For example, the onset of the strong and wide peak around 70°-120° (see Figure 4D and 8) along with the simplified drawing of Figure 2, points towards the Galactic Center with the Sun being interposed within ~5.5° around 87° (~18$^{th}$ of December).

Also here, in order to further exclude that the solar activity causes the temperature anomalies of the upper stratosphere, the attached Figure 10 shows the similarity between the simultaneously measured 10.7 cm line and the EUV emission for planetary configurations used in Figure 3-5. Then, the reasoning mentioned above for the F10.7 solar proxy applies equally to the EUV. I.e., also the solar EUV is not correlated with the annually peaking temperature distribution of the upper stratosphere. We stress this cross checking with the solar EUV irradiation because of its known direct impact on the atmosphere (above ~10 km).

In addition, Figure 8 shows an interesting comparison of the upper stratospheric temperature with that of the main ozone layer underneath (altitude ≈ 16-31 km). While the ozone layer is strongly affected by solar UV, the striking difference between both spectra in Figure 8 implies that the peaking distribution of the upper stratosphere (38.5-47.5 km) is marginally or even not at all affected by the solar UV. This striking difference of the atmospheric response between nearby layers strengthens the invisible streaming matter scenario, being eventually an important finding for the underlying particle identification (in future work).

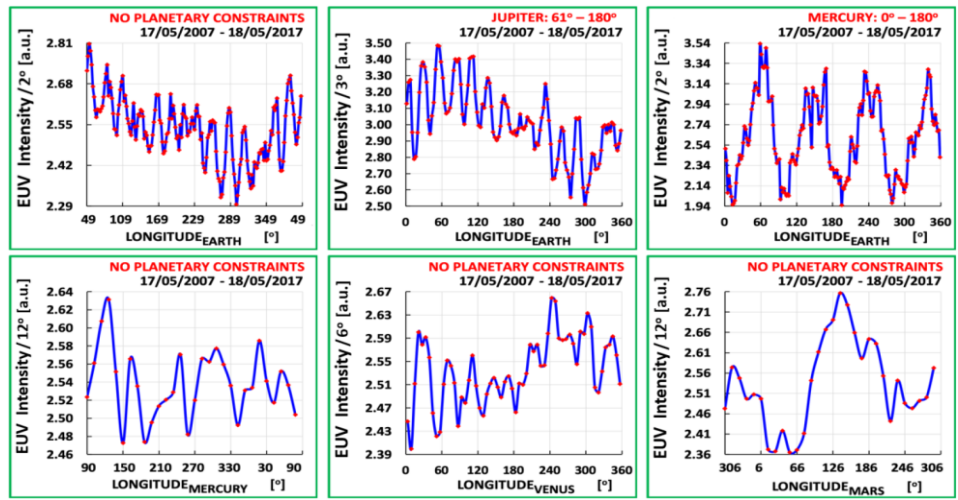

**Figure 10** For comparison with the corresponding spectra in Figure 3-5, these spectra show the solar activity associated with the daily measured EUV intensity for the same time interval (2007-2017). The similarity with the corresponding spectra of the F10.7 solar indicator is apparent (see also Figure 6(J,L)).

### 3.4 Estimation of the energy deposition

So far, for various planetary configurations, the observed dissimilarity between the measured spectral shapes of the upper stratospheric temperature and the concurrent solar activity points also at an external source, whose impact should compare with or eventually supercede the known energetic solar irradiation (above ~10-100 eV). Surprisingly, this is an enormous energy deposition for the conventional picture of a quasi not-interacting dark Universe; it excludes already that standard dark matter candidates like axions or WIMPs can be the cause, while pointing at other type of potential particle candidates (see section 4).

In the following we give an order of magnitude estimate of such an as yet overlooked energy input to the upper stratosphere, following some assumptions. The atmosphere absorbs about 15 W/m² in solar UV (~200-300 nm), which generates and interacts with the stratospheric ozone layer, and, has direct influence on the stratosphere [24,25]. Its temperature response and the UV flux modulation during the 11-years cycle is about 0.7-1.1 K and 0.2 W/m², respectively [7,25]. However, there are still uncertainties in the modelling of solar cycle impacts in the atmosphere [7,8]. In this work, the observed seasonal variation of the upper stratosphere is about ±2.5 K between solar maxima minus solar minima (Figure 11). Scaling the aforementioned values, the estimated energy deposition of the new exo-solar source should be of the order of 1 W/m². Even if this is overestimated by factor ~10 or more, it still reflects a large impact coming from the dark sector.

Our conclusion on such a macroscopic energy deposition due to some kind of invisible matter is not in conflict with the null results of the underground experiments searching for dark matter

[11]. It suffices to mention as possible reasons for this: a) Threshold effects, e.g., the stratosphere senses UV/EUV photons with an energy around 10-100 eV, and b) The stratosphere is quasi not shielded to outer space ($\rho_{overhead} \approx 1$ gr/cm$^2$), contrary to the underground dark matter experiments.

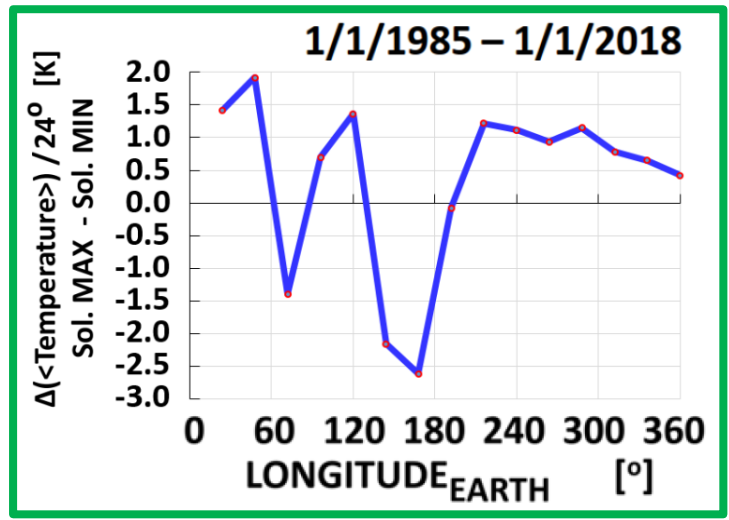

**Figure 11** Seasonal mean upper stratospheric temperature difference between periods of solar maximum and solar minimum. The absolute difference is 4.5 K. The time period used for this spectrum is 9 years for solar minimum (1985-1987, 1995-1997, 2007-2009), and, 15 years for solar maximum (1989-1993, 1999-2003, 2011-2015).

## 4. Conclusion – Summary

This work shows that "temperature anomalies", which appear every year at the upper stratosphere, can be the as yet overlooked signature of "strongly" interacting low speed streaming matter from the dark Universe. This follows mainly from the present analysis of upper stratospheric temperature measurements. The observed peaking relations exclude on their own any conventional explanation be it due to any remote planetary interaction which is smooth over a planetary orbit, or, intrinsic to the atmosphere. After various consistency tests, the main results are: a conventionally unexpected stratospheric planetary relationship suggesting an exo-solar source for the invisible streams, which give rise to a large energy deposition (~1 W/m$^2$); i.e., known energetic solar irradiation is not the only driving source behind the observed stratospheric temperature behaviour.

The implications of the driving idea of this work into the dynamic ionosphere (above ~100 km) has been considered recently [11]. In this work, we proceed on purpose towards the upper stratosphere (altitude ≈38.5-47.5 km). In so doing, we may recover in future the properties of the propagating particles by reconstructing approximately their energy deposition (in analogy to the various components of the known cosmic radiation). Though, at present, daily data on the relevant atmospheric electron content (TECUs) refers to the entire ionosphere. Interestingly, the Earth's ionosphere has an anomalously high degree of ionization in December, which is known since 1937 [30]. Remarkably, this coincides with the annual alignment between Earth (Moon), Sun and the Galactic Center, and, with the striking peaking stratospheric temperature anomalies derived here (see e.g. Figure 8). Hence, once spatially accurate ionospheric data become available, it will be promising to search for possible correlations between the dynamic behaviour of the stratosphere and the ionosphere above.

Finally, it is worth mentioning that contrary to the wide belief of an extremely feebly interacting dark Universe, also recent work discusses potential dark matter constituents with a rest mass around ~$10^{-3}$ GeV/$c^2$ [12], and, ~$10^{+25}$ GeV/$c^2$ [13,14], having a very large cross section with normal matter. Both suggestions fit-in the aforementioned observationally derived conclusion of invisible matter with a large cross section with normal matter without contradicting cosmological arguments (2.7 K CMB radiation). Therefore, they deserve further attention.

Future observations may extend this first analysis to the whole atmosphere, including larger altitudes; newly released datasets cover 80 km in altitude with 60 isobaric layers [19]. Analyzing more atmospheric observations, the nature of the assumed "invisible matter" could be deciphered.

**Data availability**

The data that support the findings of this study are available from: **a)** the Planetary Ephemerides for the various daily planetary positions have been taken from NASA JPL Horizons system (https://ssd.jpl.nasa.gov/horizons.cgi), **b)** the daily observations of solar EUV have been taken from Solar EUV Monitor (SEM) on the solar Heliospheric Observatory (SOHO) (https://dornsifecms.usc.edu/space-sciences-center/download-sem-data/) (Leonid Didkovsky), **c)** the daily F10.7 measurements are provided courtesy of the National Research Council Canada in partnership with the Natural Resources Canada (http://www.spaceweather.ca/solarflux/sx-5-en.php) but they have been obtained from the GSFC/SPDF OMNIWeb interface at https://omniweb.gsfc.nasa.gov where F10.7 is adjusted for 1 A.U, (for middle hour: 20:00), and **d)** the stratospheric temperatures have been downloaded from European Centre for Medium-range Weather Forecast (ECMWF) available from https://www.ecmwf.int/en/forecasts/datasets/archive-datasets/reanalysis-datasets/era-interim .

**ACKNOWLEDGMENTS**

We wish to sincerely thank the anonymous referee for the feedback, which helped us to improve by a lot the initial manuscript. The work by Y.K.S was supported in part by IBS-R017-D1-2019-a00. For M.M., this research is co-financed by Greece and the European Union (European Social Fund- ESF) through the Operational Programme «Human Resources Development, Education and Lifelong Learning» in the context of the project "Strengthening Human Resources Research Potential via Doctorate Research" (MIS-5000432), implemented by the State Scholarships Foundation (IKY).

**REFERENCES:**

[1] V. F. Hess, Über Beobachtungen der durchdringenden Strahlung bei sieben Freiballonfahrten, Physikalische Zeitschrift 13 (**1912**) 1084; see also arXiv:1808.02927v2 translated, commented by A. De Angelis, C. Arcaro b. Schultz.

[2] Q. Zhang, C.-S. Shin, H. van den Dool, M. Cai, CFSv2 prediction skill of stratospheric temperature anomalies, Clim. Dyn. 41 (**2013**) 2231; https://doi.org/10.1007/s00382-013-1907-5 .

[3] H.-T. Zhou, A.J. Miller, J. Wang, J.K. Angel, Downward-Propagating Temperature Anomalies in the Preconditioned Polar Stratosphere, J. Climate 15 (**2002**) 781; https://doi.org/10.1175/1520-0442(2002)015<0781:DPTAIT>2.0.CO;2.

[4] R. Thiéblemont, K. Matthes, N. E. Omrani, K. Kodera, F. Hansen, Solar forcing synchronizes decadal North Atlantic climate variability. Nat. Commun. 6 (**2015**) 8268; https://doi.org/10.1038/ncomms9268 .

[5] K. Matthes, Y. Kuroda, K. Kodera, U. Langematz, Transfer of the solar signal from the stratosphere to the troposphere: Northern winter, J. Geophys. Res. 111 (**2006**) D06108, and ref's therein; https://doi.org/10.1029/2005JD006283 .

[6] J.P. McCormack, L.L. Hood, Apparent solar cycle variations of upper stratospheric ozone and temperature: latitudinal and seasonal dependences, J. Geophys. Res., 101 (**1996**) 20933; https://doi.org/10.1029/96JD01817 .

[7] E.M. Bednarz, A.C. Maycock, P.J. Telford, P. Braesicke, N. L. Abraham, J.A. Pyle, Simulating the atmospheric response to the 11-year solar cycle forcing with the UM-UKCA model: the role of detection method and natural variability, Atmos. Chem. Phys. 19 (**2019**) 5209; https://doi.org/10.5194/acp-19-5209-2019

[8] S. S. Dhomse, M. P. Chipperfield, R. P Damadeo, J. M. Zawodny, W. T. Ball, W. Feng, R. Hossaini, G.W. Mann, J.D. Haigh, On the ambiguous nature of the 11 year solar cycle signal in upper stratospheric ozone, Geophys. Res. Lett., 43 (**2016**) 7241; https://doi.org/10.1002/2016GL069958 .

[9] D.H. Hathaway, The Solar Cycle, Living Rev. Solar Phys. 12 (**2015**) 4; http://dx.doi.org/10.1007/lrsp-2015-4 .

[10] K. Zioutas, M. Tsagri, Y.K. Semertzidis, D.H.H. Hoffmann, T. Papaevangelou, V. Anastassopoulos, The 11 years solar cycle as the manifestation of the dark Universe. Mod. Phys. Lett. A29 (**2014**) 1440008; https://doi.org/10.1142/S0217732314400082 (https://arxiv.org/abs/1309.4021 ).

[11] S. Bertolucci, K. Zioutas, S. Hofmann, M. Maroudas, The Sun and its Planets as Detectors for invisible matter, Physics Dark Universe 17 (**2017**) 13, and ref's therein; https://doi.org/10.1016/j.dark.2017.06.001.

[12] T. Emken, R. Essig, C. Kouvaris, M. Sholapurkar, Direct Detection of Strongly Interacting Sub-GeV Dark Matter via Electron Recoils, CERN-TH-2019-071, CP3-Origins-2019-18 DNRF90, YITP-SB-19-14, e-print https://arxiv.org/abs/1905.06348 (**2019**).

[13] A.R. Zhitnitsky, "Nonbaryonic" Dark Matter as Baryonic Color Superconductor, J. Cosmol. Astropart. Phys., 0310 (**2003**) 010; https://arxiv.org/abs/hep-ph/0202161 .

[14] N. Raza, L. Van Waerbeke, A. Zhitnitsky, Solar corona heating by axion quark nugget dark matter, Phys. Rev. D98 (**2018**) 103527; https://doi.org/10.1103/PhysRevD.98.103527 .

[15] K. Lawson, A.R. Zhitnitsky, The 21 cm absorption line and the axion quark nugget dark matter model, Physics Dark Universe 24 (**2019**) 100295; https://doi.org/10.1016/j.dark.2019.100295.

[16] D. P. Dee, S. M. Uppala, A. J. Simmons, P. Berrisford, P. Poli, S. Kobayashi, U. Andrae, M. A. Balmaseda, G. Balsamo, P. Bauer, P. Bechtold, A. C. M. Beljaars, L. van de Berg, J. Bidlot, N. Bormann, C. Delsol, R. Dragani, M. Fuentes, A. J. Geer, L. Haimberger, S. B. Healy, H. Hersbach, E. V. Hólm, L. Isaksen, P. Kållberg, M. Köhler, M. Matricardi, A. P. McNally, B. M. Monge-Sanz, J.-J. Morcrette, B.-K. Park, C. Peubey, P. de Rosnay, C. Tavolato, J.-N. Thépaut, F. Vitart, The ERA-Interim reanalysis: configuration and performance of the data assimilation system, Q. J. Royal Meteorol. Soc. 137 (**2011**) 553; https://doi.org/10.1002/qj.828.

[17] P. Berrisford, D. P. Dee, P. Poli, R. Brugge, K. Fielding, M. Fuentes, P. W. Kållberg, S. Kobayashi, S. Uppala, A. Simmons, The ERA-Interim archive Version 2.0. ERA Report Series 1 (**2011**); http://www.ecmwf.int/en/elibrary/8174-era-interim-archive-version-20 .

[18] European Centre for Medium-range Weather Forecast (ECMWF) (**2011**) The ERA-Interim reanalysis dataset, Copernicus Climate Change Service (C3S) (accessed Dec 2018, Jan **2019**), available from https://www.ecmwf.int/en/forecasts/datasets/archive-datasets/reanalysis-datasets/era-interim .

[19] Release of ERA5 reanalysis data (1979-):http://climate.copernicus.eu/climate-reanalysis. See also http://climate.copernicus.eu/c3s-user-service-desk  (**2019**).

[20] M. Agostini et al., The Borexino collaboration, Modulations of the cosmic muon signal in ten years of Borexino data, JCAP 1902 (**2019**) 046; https://doi.org/10.1088/1475-7516/2019/02/046  http://arxiv.org/abs/1808.04207 .

[21] NOAA, National Weather Service, CPC, Stratosphere: Global Temperature Time Series, https://www.cpc.ncep.noaa.gov/products/stratosphere/temperature/  (**2018**).

[22]  E. Möbius, P. Bochsler, M. Bzowski, D. Heirtzler, M.A. Kubiak, H.  Kucharek, M.A. Lee, T. Leonard, N.A. Schwadron, X. Wu, S.A. Fuselier, G. Crew, D.J. McComas, L. Petersen, L. Saul, D. Valovcin, R. Vanderspek, P. Wurz, Interstellar Gas Flow Parameters Derived From Interstellar Boundary Explorer-Lo.Observations In 2009-2010: Analytical Analysis, Astrophys. J. Suppl., 198 (**2012**) 11, see Figure 2 therein; http://iopscience.iop.org/0067-0049/198/2/11/ .

[23] H. Fischer, X. Liang, Y. Semertzidis, A. Zhitnitsky, K. Zioutas,  New mechanism producing axions in the AQN model and how the CAST can discover them, Phys. Rev. D98 (**2018**) 043013, see relation (13), section V. A. and ref's 49-53 therein; https://doi.org/10.1103/PhysRevD.98.043013.

[24] J. Lean, G. Rottman, J. Harder, G. Kopp, SORCE Contributions to new understanding of global change and solar variability,  Solar Physics 230 (**2005**) 27; https://doi.org/10.1007/0-387-37625-9_3  p.32.

[25] D. Rind, J. Lean, J. Lerner, P. Lonergan, A. Leboissitier, Exploring the stratospheric / tropospheric response to solar forcing, J. Geophys. Res., 113 (**2008**) D24103; https://agupubs.onlinelibrary.wiley.com/doi/full/10.1029/2008JD010114 .

 [26]  A.H. Butler, J.P. Sjoberg, D.J. Seidel, K.H. Rosenlof, A sudden stratospheric warming compendium, Earth Syst. Sci. Data 9 (**2017**) 63;https://doi.org/10.5194/essd-9-63-2017.

[27] S.A. Sitnov, QBO effects manifesting in ozone, temperature, and wind profiles, Annales Geophysicae 22 (**2004**) 1;  http://ifaran.ru/troica/biblio/sitnov/PDF3.pdf  .

[28] M. P. Baldwin  L. J. Gray  T. J. Dunkerton  K. Hamilton  P. H. Haynes  W. J. Randel  J. R. Holton  M. J. Alexander  I. Hirota  T. Horinouchi  D. B. A. Jones  J. S. Kinnersley, C. Marquardt  K. Sato  M. Takahashi, The quasi-biennial oscillation, Reviews of Geophysics 39 (**2001**) 179; https://doi.org/10.1029/1999RG000073 .

[29] C. H. O'Reilly, A. Weisheimer, T. Woollings, L. J. Gray, D. MacLeod, The importance of stratospheric initial conditions for winter North Atlantic Oscillation predictability and  implications for the signal-to-noise paradox, QJRMS 145 (**2019**) 131; https://doi.org/10.1002/qj.3413 .

**[30]** E.V. Appleton, Regularities and irregularities in the ionosphere, Proc. Royal Soc. of London A162 (**1937**) 451; https://royalsocietypublishing.org/doi/abs/10.1098/rspa.1937.0195 .